# Plate-like precipitate effects on plasticity of Al-Cu alloys at micrometer to sub-micrometer scales

Peng Zhang[a], Jian-Jun Bian[a], Chong Yang[a], Jin-Yu Zhang[a], Gang Liu[a], Jérôme Weiss[b], Jun Sun[a]

[a]State Key Laboratory for Mechanical Behavior of Materials, Xi'an Jiaotong University, Xi'an, 710049, China
[b]IsTerre, CNRS/Université Grenoble Alpes, 38401 Grenoble, France

**Abstract**─The continuous miniaturization of modern electromechanical systems calls for a comprehensive understanding of the mechanical properties of metallic materials specific to micrometer and sub-micrometer scales. At these scales, the nature of dislocation-mediated plasticity changes radically: sub-micrometer metallic samples exhibit high yield strengths, however accompanied by detrimental intermittent strain fluctuations compromising forming processes and endangering structural stability. In this paper, we studied the effects of plate-like $\theta'$-Al$_2$Cu precipitates on the strength, plastic fluctuations and deformation mechanisms of Al-Cu alloys from micro-pillar compression testing. The plate-like precipitates have diameters commensurate with the external size of the Al-Cu micro-pillars. Our results show that these plate-like precipitates can strengthen the materials and suppress plastic fluctuations efficiently at large sample sizes ($\gtrsim 3\ \mu m$). However, the breakdown of the mean-field pinning landscape at smaller scales weakens its taming effect on intermittency. Over an intermediate range of sample sizes allowing the precipitates to cross the entire pillar, an enhanced "apparent strain hardening" and a sharp decrease of jerkiness are observed, in association with the presence of {100}-slip traces along the coherent $\theta'$-Al$_2$Cu precipitate/$\alpha$-Al matrix interface and precipitate shearing. These complex effects of plate-like precipitates on plasticity are analyzed, experimentally and theoretically, in view of the interferences between external and internal sizes, and the related modifications of the underlying plastic mechanisms.

# 1. Introduction

Over recent years, the structural analysis and design of metallic structures in the field of micro- and nano-electromechanical systems (MEMS and NEMS) highlighted the need to investigate the specificity of the mechanical properties of metallic materials at those scales [1-3]. The vast amount of work on single crystals published during the last decade [4-8] revealed several size effects characterizing micro-plasticity: (i) a "smaller is stronger" effect, caused by novel plastic mechanisms commonly referred to "source truncation" [9-12] and "dislocation starvation" [13,14]; (ii) a modification of hardening mechanisms, from classical forest-hardening at large length scales to exhaustion hardening [9,15,16]: to sustain the flow in the case where weak dislocation sources are exhausted, much higher stress is required for the activation of the stronger, truncated ones. These studies depicted a different landscape, away from conventional engineering, and paved the way towards size-dependent design [1].

Besides an enhanced plastic yield stress, another specific feature of sub-$\mu m$ plasticity is the intermittent character of plastic flow, which manifests itself directly on the stress-strain curves through discrete strain bursts with a broad range of sizes following power-law tailed distributions [17-21]. These collective effects and dislocation avalanches are rooted in the long ranged elastic interactions between dislocations, hence challenging some fundamental aspects of classical plasticity theory based on the behavior of individual dislocations, or which is assuming an average response of dislocation ensembles [7,22-24]. They have been regarded as a common feature of small sized single crystals, prevailing not only in face-centered cubic (FCC), but also in body-centered cubic (BCC) and hexagonal closed-packed (HCP) structures [25-28]. In contrast, in bulk materials, the dominating role of power-law distributed fluctuations has only been identified in HCP crystals characterized by single slip plasticity [29-33]. In FCC bulk materials characterized by multislip plasticity, short-range interactions prevail and dislocation networks develop, giving rise to numerous, small and uncorrelated fluctuations following a Gaussian distribution, while the rare remaining power-law distributed fluc-tuations are associated with sudden rearrangements of the dislocation substructures [32]. This difference between small scale and bulk plasticity implies an external size effect on plastic intermittency.

Recently, the present authors performed detailed analyses of the intermittent character of plasticity in compressed micro-pillars of Al and Al alloys, and defined the wildness $W$ as the fraction of plastic deformation accommodated through power-law distributed fluctuations (wild plasticity) [17]. These experiments demonstrated that diminishing the external length scale intensifies fluctuations (higher degree of wildness, $W$), shifting the collective dislocation dynamics towards criticality, with a progressively smaller exponent $\kappa$ for the power law-tailed distribution of burst sizes $X$, $P(X) \sim X^{-\kappa}$. From the engineering viewpoint, the external size related plastic fluctuations are unwelcome due to their unpredictable nature and the possibility for large avalanches to span the whole system size [19,34]. Giving a practical example, the mass production of metallic patterns with high resolution and quality (e.g., ultra-smooth surfaces) represents a substantial challenge, partly limited by undesired morphological fluctuations resulting from stochastic dislocation avalanches [35]. Thus, developing methodologies to tame the detrimental effects of plastic fluctuations in a predictable and quantitative manner is of critical importance for the structural analysis and design of metallic structures at small scales [19].

Some previous investigations [36-39] suggested that smooth plastic flow can be achieved via the introduction of quenched disorder, thereby providing the potential for inhibiting intermittency. This motivated us to thoroughly explore this strategy in Al alloys micro-pillars containing either solutes or precipitates. We demonstrated that alloying frustrates the propagation of dislocation avalanches, hence shifts the transition from "mild" to "wild" plasticity towards smaller external size and increases the power-law exponent $\kappa$ of the avalanche distribution (i.e. a relatively lower probability for large avalanches) [17]. To quantify the competition between external size and disorder in one framework, we considered an internal length $l = Gb/\tau_{pin}$, where $\tau_{pin}$ is the pinning strength to dislocation motion, originating in alloys mainly from disorder arrays, $G$ is the shear modulus and $b$ is the magnitude of Burger vector. We found that a single dimensionless parameter $R = L/l$ controls the wild-to-mild transition and the scaling exponent $\kappa$ for all the materials tested, where $L$ is the system size (micro-pillar diameter). Actually, the defined internal length scale $l$ corresponds to a distance at which the long-range elastic interaction stress between dislocations becomes comparable with the pinning strength of the obstacles [22]. This central role of $R = L/l$ in different materials shows that the effect of disorder on intermittency occurs through a thwarting of long-range interactions.

However, these external size and disorder effects on plastic intermittency were only quantified for [110]-oriented Al alloys containing nano-sized disorders [17], i.e. FCC metallic single crystals with multislip orientation and microstructural disorder length scales (individual disorder size and disorder spacing) much smaller than the external size. However, the advantages of using quenched disorders to tame plastic intermittency, compared with other effective methodologies such as grain boundary engineering and coating [39-41], not only come from the fact that precipitation processes (aging treatment) are well controlled [42], but also from the variety of disorder-reinforced alloying systems that can be selected. In many industrial alloys, such as the 2XXX series Al alloys [43] and Ni-Al based alloys [44], the size of precipitates can be of the order of few micrometers, i.e. comparable to, or even larger than, the dimension of small technological devices. Consequently, we extend here our former work [17] to alloys containing large-sized disorders, examine to what extent the interactions between internal and external length scales fit the previously proposed theoretical framework [17,32], and show how to take advantage of such type of disorder.

In this study, we systematically explored strength and plastic fluctuations in small-sized Al-2.5wt% Cu and Al-4.0wt% Cu single crystals containing plate-like $\theta'$-Al$_2$Cu precipitates. The average precipitate diameter is about 1 $\mu m$, comparable with the inter-particle spacing and the sample size (300 to 4000 nm). Overall, we verified that the external and internal size effects on strengthening as well as on the wild-to-mild transition can fit into the previous frameworks. However, the strong interferences between the external size $L$ and the microstructural disorder length scales results in a complex $L$-dependent plasticity. A sharp decrease of wildness, which is associated with the presence of {100}-slip traces and precipitate shearing, was observed over a specific diameter range. We rationalized these phenomena by a combination of length scale analyses and microstructure observations. The underlying mechanism giving rise to the non-trivial {100} slip traces is analyzed in a companion paper [45].

## 2. Experimental procedures

Two kinds of Al-Cu alloys, one with a composition of Al-2.5wt.% Cu (abbreviated Al-2.5Cu) and the other with Al-4.0wt.% Cu (Al-4.0Cu), were studied. The alloys were melted and cast in a stream argon, by using 99.99 wt.% pure Al and Al-50 wt.% Cu master-alloy, followed by a homogenization at 723 K for 24 h to eliminate composition segregation. Specimens with cross-section of 10 mm × 10 mm were cut from the central region of the ingots and subjected to further heat treatments. All alloys were solutionized at 793 K for 4 h, followed by a cold-water quench and immediately aged at 523 K for 8 h. The maximum error of all the temperature measurements in the present experiments was ±1 K. Precipitates in the aged alloys were quantitatively characterized by standard transmission electron microscopy (TEM) (Fig. 1a and b) and high resolution transmission electron microscopy (HRTEM). Details about these microstructural measurements can be found in previous publications [17,41,46].

An electron backscatter diffraction (EBSD) detector integrated inside the focus ion beam (FIB) chamber was used to probe the crystal orientation on the electro-polished surfaces of the aged samples (Fig. 1d). The grains with their normal along the [110]-direction, an orientation identical to the Al-alloys micro-pillars we studied before [17], were selected to fabricate micro-pillars by FIB micro-machining. As the grain size is of the order of few mm in the present cast metals (Fig. 1d), a large number of micro-pillars can be fabricated within a single [110]-grain. Micro-pillars were prepared with a diameter ranging from 300 to 4000 nm at mid sample height, with an aspect ratio of ~ 3:1 (see Fig. 1e and f).

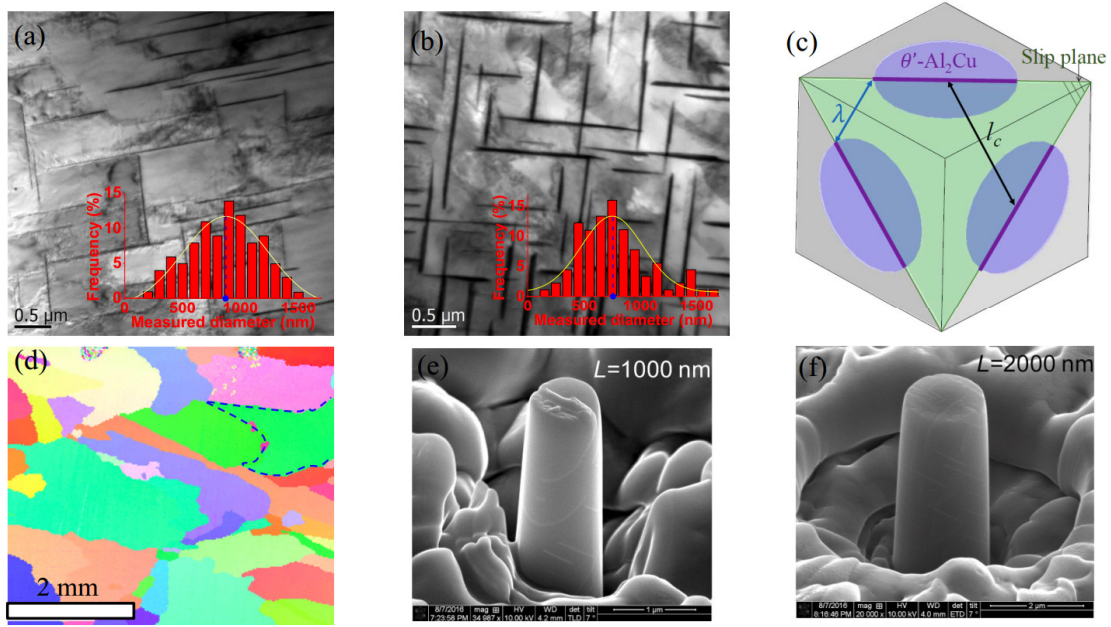

**Figure 1. The microstructure of tested materials and fabricated micro-pillars.** **(a)** and **(b)** TEM images of Al-2.5Cu and Al-4.0Cu alloy respectively, taken along the {100}-direction. The insets in (a) and (b) show the distributions of measured precipitate diameters **(c)** Schematic diagram showing an ideal distribution of $\theta'$-precipitates in the matrix. **(d)** The EBSD mapping, where the [110]-orientated grain is marked by dot dash lines. **(e)** and **(f)** Typical images of Al-4.0Cu micro-pillars before compression, with 1000 and 2000 nm diameter respectively. The traces of intersections between the precipitates and the pillar surface can be observed.

The ex-situ micro-compression tests were performed on a nano-indentation system (Hysitron Ti 950) with a 10 μm flat diamond tip. Following our previous test conditions [17], the micro-pillars were compressed at room temperature under a displacement-controlled mode, with a strain rate of ~2×10$^{-4}$ s$^{-1}$ up to 18% engineering strain. The engineering stress-strain curves were obtained by using the diameter at half height of micro-pillars and the initial height. In order to explore the effects of precipitates on sub-μm plasticity in more details, in-situ TEM compression testing was also performed on [110]-oriented Al-2.5Cu alloy micro-pillars, by using a nano-indentation system (Hysitron Pi 95) inside the TEM chamber. Further details on micro-pillar fabrication and this mechanical testing methodology can be found in Ref. [47-49]. In addition, to evaluate the total strengthening contribution from obstacles, tensile tests were performed on bulk polycrystalline Al-Cu by using a servohydraulic Instron-1195 testing machine, with a same strain rate as in micro-compression tests. The tensile specimens were dog-bone shaped, with a gauge size of 15 mm in diameter and a length of 75 mm.

# 3. Characterization of tested materials

In this section, we characterize the tested materials in terms of characteristic length scales and pinning strength of precipitate arrays, which are crucial to understand the experimental results.

## *3.1 Characteristic length scales of oriented precipitates*

The TEM images show a large number of $\theta'$-Al$_2$Cu precipitates formed on the $\{100\}_\alpha$ plane, with a number density greater in the Al-4.0Cu alloy (Fig. 1b) than in the Al-2.5Cu alloy (Fig. 1a). The $\theta'$-Al$_2$Cu precipitates have a plate-shaped morphology with coherent $(100)_{\theta'}\|(100)_{\alpha\text{-Al}}$ interfaces along the broad face and semi-coherent interfaces around the rim of the plates [42]. Previous reports stated that such structured $\theta'$-Al$_2$Cu phase is non-shearable to dislocations under engineering tensile testing [50-53].

The characteristic length scales of precipitate arrays include their average size and spacing. The precipitate size can be directly measured from the TEM images. These measured precipitate sizes are broadly distributed (insets of Fig. 1a and b). However, it should be mentioned that TEM measurements inevitably exaggerate this scattering to some extent for stereological reasons, as the TEM foils, with a thickness (measured as ~ 70 nm) much smaller than the precipitate diameter, truncate the precipitates at a random position. Similarly, the real average precipitate diameter should be larger than the measured one, and can be corrected as follows [52]:

$$d_d = \frac{2}{\pi}\left(d_m - t_f + \sqrt{(d_m - t_f)^2 + 4\pi d_m t_f}\right) \qquad (1)$$

where $d_m$ and $d_d$ are measured and real mean diameters, respectively, and $t_f$ is the TEM foil thickness. The corrected average diameters $d_d$ are comparable in Al-2.5Cu and Al-4.0Cu alloys (~ 1 μm, see table 1).

To extract the average spacing between oriented precipitates, we consider an ideal distribution landscape of $\theta'$-Al$_2$Cu precipitates [42,52], assuming that all plate-shaped precipitates are located at the center of each (100) surface of a cubic matrix volume (Fig. 1c). In this case, the precipitates will form a triangular array in the section of {111}-slip planes, as schematically shown in Fig. 1c. The average nearest spacing between precipitates, $\lambda$, is of particular importance, as it determines the critical stress for dislocations to overcome precipitates, hence the yield strength. For point-like disorder arrays ($\lambda \gg d_d$), this parameter $\lambda$ solely characterizes the precipitate spacing. The situation is different here. In particular, one can keep $\lambda$ unchanged while constructing a broader spacing inside the triangular array by enlarging the precipitate diameter $d_d$. Consequently, another important length scale of precipitate spacing, the center-to-center distance, $l_c$, should be introduced (Fig. 1c) [52]. It characterizes the dimension of the space surrounding by plate-like precipitates, and represents the distance that the dislocation segments can cross before encountering precipitates.

Using the experimentally measured precipitate parameters as summarized in Table 1, the spacing parameters $\lambda$ and $l_c$ can be determined from [52]:

$$\lambda = \frac{1.2669}{\sqrt{N_v d_d}} - \frac{\pi d_d}{8} - 1.061 t_d \tag{2}$$

$$l_c = \lambda + \frac{d_d}{2} + \frac{\sqrt{3}}{2} t_d \tag{3}$$

where $N_v$ is the precipitate number density, and $t_d$ the precipitate thickness. The values of $\lambda$ and $l_c$ are presented in Table 1.

**Table 1. Statistical and calculated results on the microstructure in the Al-Cu alloys.**

| Materials | Precipitate diameter, $d_d$ (nm) | Precipitate thickness, $t_d$ (nm) | Number Density, $N_v$ ($10^{18}$ m$^{-3}$) | Volume fraction (%) | Pinning strength (MPa) | Nearest precipitate spacing $\lambda$ (nm) | *Center-to-center spacing $l_c$* (nm) | Internal scale *l* (nm) |
|---|---|---|---|---|---|---|---|---|
| **Al-2.5Cu** | 1190 (67) | 13.0 (1.8) | 1.69 (0.08) | 2.44 (0.12) | 35.0 (2.4) | 412 (38) | 1007 (82) | 197 (14) |
| **Al-4.0Cu** | 973 (85) | 14.3 (2.9) | 4.66 (0.30) | 4.92 (0.38) | 47.4 (2.1) | 192 (17) | 678 (47) | 149 (7) |

The values in the brackets are the measurement errors of microstructural parameters and bulk tension tests (pinning strength).

*3.2 Evaluation of Pinning strength*

The pinning strength, i.e. the resistance to dislocation motion arising from lattice resistance, forest dislocations, and quenched disorders, has been proved to be a critical parameter in controlling dislocation avalanches [17,38]. Following our previous work, we evaluate this pinning strength from the experimental measurements of yield strength, $\sigma_y$, of the bulk Al-2.5Cu and Al-4.0Cu polycrystalline alloys by tensile testing to avoid the external size effect on strength [17]. Because the studied bulk Al-Cu alloys have a texture-free, equiaxial grain structure, the pinning strength can be estimated as follows [54]:

$$\tau_{pin}^{bulk} = \sigma_y/M - kL_g^{-1/2} \qquad (4)$$

where $M$ is the Taylor factor (3.06 for FCC), $L_g$ the average grain size, and $k$ the Hall-Petch constant (60 MPa·µm$^{1/2}$ [55]). The yield strength, $\sigma_y$, is measured as the 0.2% offset stress on the stress-strain curves of the bulk material. Note that the grain boundary strengthening is almost negligible compared with pinning strength, owing to the very large grain size in these alloys (~ 1mm, Fig. 1d). The evaluated pinning strength of Al-2.5Cu alloy is ~35 MPa, smaller than ~47 MPa in Al-4.0Cu alloy (Table 1). Note that the so-evaluated $\tau_{pin}^{bulk}$ can only be used in micro-pillars with bulk-like dislocation-precipitate interaction, i.e., Orowan mechanism on {111}-plane, but becomes irrelevant if other interaction mechanisms take place.

## 4. Results

### *4.1 Plastic deformation curves of micro-pillars*

The shortcomings of the micro-pillar compression methodology, such as instrumental constraints [56], sample imperfections [47] and substrate deformation [49], can lead to non-uniform stress and strain distributions, making the true stress and strain largely inaccessible, especially under large deformation. A detailed discussion on the testing methodology and its influence on the mechanical characterization has been proposed recently [57]. In view of these drawbacks, we simply present here engineering stresses and strains, and only use the stress values at small strain (0.2% plastic strain and 2% total strain) in our discussion on micro-pillar strength, in order to avoid the increasing errors caused by the lateral constraints and possible strain localization as deformation goes on.

Some representative engineering shear stress-strain curves are presented on Fig. 2 to highlight the external size and internal disorder effects, and their interplay, on plastic flow. At large external size $L \geq 3500$ nm, both alloys deform in a smooth and continuous manner, similarly to their bulk counterparts, but with a higher yield strength (the stress at 0.2% plastic strain) (Fig. 2d). Decreasing the sample size to $L \approx 2000$ nm, the plastic deformation of the Al-2.5Cu alloy becomes jerky, characterized by pronounced strain bursts and stress drops scattered on the stress-strain curves (Fig. 2c). Such interrupted plastic flow is the signature of dislocation avalanches under strain-rate controlled compression: the hysteresis of instrumental autoregulation (PID feedback) during a rapid release of plastic strain implies that strain bursts are accompanied by abrupt stress drops triggered by a retraction of the compression tip to its setting position [13,58]. The disorder effect can be identified in Fig. 2c, where bursts appear smaller and rarer in Al-4.0Cu alloy. Decreasing further the external size to $L = 500$ nm, strain bursts become more drastic, and of similar size in the two alloys (Fig. 2a).

Besides these common features in the disorder-reinforced small samples [8,36,37], two anomalous phenomena can be identified in these Al-Cu alloys. First, we find that the deformation curves of Al-Cu micro-pillars seem to be jerkier than for micro-pillars containing nano-sized disorders. As an example, the 1800 nm sized micro-pillar of Al-2.5Cu (Fig. 2c) is much jerkier than the "Al-Sc precipitate" alloy with similar size we studied before (Fig. 3a in Ref. [17]), despite a similar pinning strength for the two alloys (~35 MPa). This point will be discussed in section 5.3.1. Second, it can be observed that the size of load drops in 700 nm Al-2.5Cu micro-pillar (Fig. 2b) seems smaller (less jerky) than in 1800 nm sample (Fig. 2c), in contradiction with the "smaller is wilder" effect. This phenomenon will be explored in more details in sections 4.4.2 and 5.3.2.

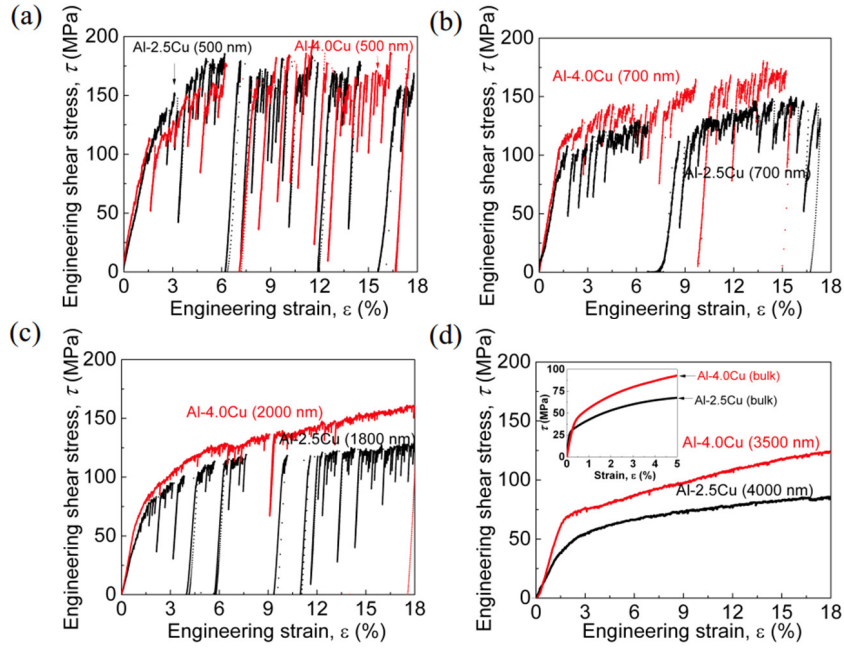

**Figure 2. Representative engineering shear stress-strain curves.** Engineering shear stress-strain curves of Al-2.5Cu and Al-4.0Cu alloys with diameter of **(a)** 500 nm, **(b)** 700 nm, **(c)** 1800 and 2000 nm respectively**,** and **(d)** 4000 and 3500 nm respectively. The shear stress $\tau$ is calculated by multiplying the compression stress with Schmid factor on {111}-plane. The shear stress-strain curves for bulk polycrystalline samples are given in the inset of (d).

*4.2 Deformation morphology and its correlation with deformation curves*

Representative SEM images of compressed Al-Cu micro-pillars with different sample sizes are shown on Fig. 3 and Fig. 4. Large specimens of both Al-2.5Cu and Al-4.0Cu alloys display numerous fine {111}-slip traces terminating inside the micro-pillars, and some of them intersecting with each other, as marked in Fig. 3a and c by white arrows ($L$ = 4000 and 3500 nm respectively). These features can be linked to the inhibition of slip propagation across the entire sample by precipitates and dislocation short-range interactions, both of which are tightly link to mild plasticity [17,32]. Indeed, smooth deformation curves (Fig. 2d) and large mildness (small $W$, section 4.4) are observed in these samples. Diminishing sample size leads to a gradual increase of plastic anisotropy, evidenced by the coarse slip bands observed in Fig. 3b and d ($L$ = 1800 and 2000 nm respectively). This external size effect on plastic anisotropy and strain localization can be understood by a limited number of available sources lying on a few slip planes [9,10], and by the reduction of dislocation cross-slip, multiplication and storage before annihilation at free surface [13,59]. In this case, the long-range elastic interactions dominate the collective behavior of dislocations, resulting in the observed jerky stress-strain curves in Fig. 2c. The deformation curves of Al-4.0Cu alloy are less jerky than Al-2.5Cu alloy in Fig. 2c, consistent with a relatively less concentrated deformation and more slip bands when comparing Fig. 3d with Fig. 3b.

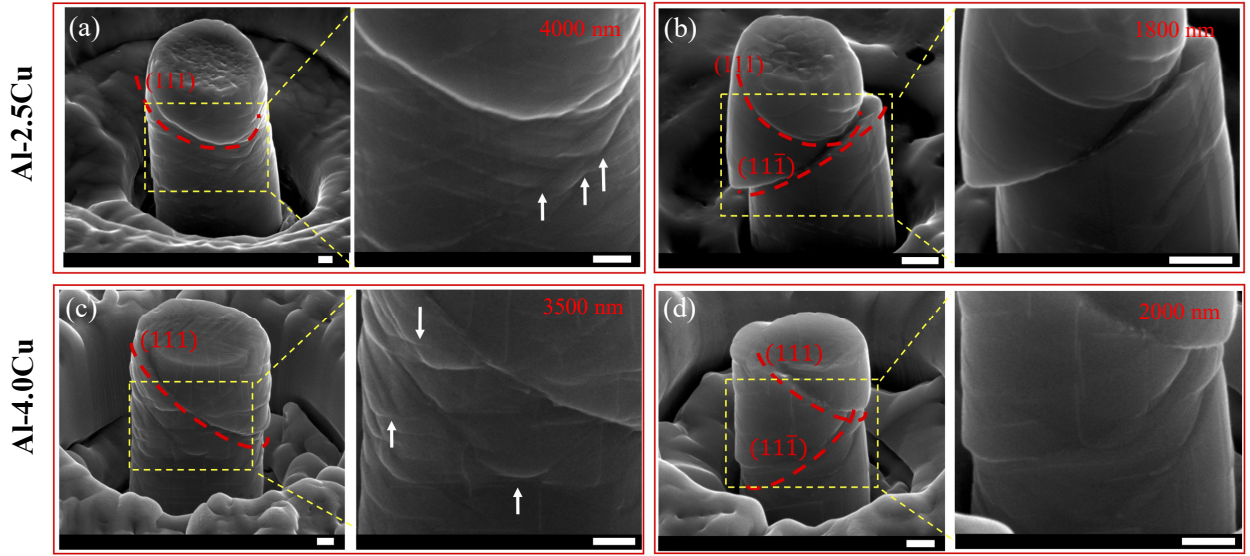

**Figure 3. Representative SEM images showing the deformation morphology of Al-Cu alloys after 18% compression strain.** Al-2.5Cu micro-pillars with diameters $L$ = 4000 nm **(a)**, 1800 nm **(b)**, and Al-4.0Cu micro-pillars with diameters $L$ = 3500 nm **(c)**, 2000 nm **(d)**. The zoom views of the images show the slip traces more clearly. The white arrows in (a) and (b) mark the intersections of slip traces. The Miller indices of coarse slip traces are marked according to the sketches shown in Fig. 4. All the scale bars represent 500 nm.

On the other hand, when the external size matched the precipitate diameter, $L/d_d \approx 1$, we observed in some micro-pillars highly distorted deformation morphologies, showing not only traditional {111}-slip traces, but also slip traces along non-close-packed planes (Fig. 4a and d). This unexpected phenomenon was only observed over a limited diameter range, 700≤ $L$ ≤1000 nm for Al-2.5Cu and 500≤ $L$ ≤1000 nm for Al-4.0Cu alloy, and disappeared upon further decreasing the sample size (Fig. 4b). The trace analysis (Fig. 4c and f) suggests that these non-trivial slip traces are caused by slip along {100}-planes. According to our knowledge, {111}-slip has been considered as the only slip mode at room temperature in pure Al and Al alloys, even in the nano-sized samples under an external stress approaching the theoretical strength [12,60]. In a recent work [45], we show that this {100}-slip occurs along the coherent $\theta'$-Al$_2$Cu precipitate/$\alpha$-Al matrix interface, and we analyze the mechanisms giving rise to this unusual interfacial slip. Here we focus on the consequences of this slip mode in terms of mechanical behavior and plastic intermittency.

To illustrate the role of {100}-slip on the mechanical behavior, we present two compressed Al-4.0Cu micro-pillars of the same size ($L = 1000$ nm) in Fig. 4d and e: the first one displays a distorted morphology and obvious {100}-slip traces (Fig. 4d), while the other one exhibits coarse slip traces along {111}-plane (Fig. 4e). The corresponding engineering stress-strain curves are given in Fig. 4g and h. In micro-pillars with {100}-slip traces, the initial jerky behavior progressively vanishes with increasing strain while some "apparent strain hardening" (engineering stress increase with strain) is observed, in strong contrast with micro-pillars showing {111}-slip only. As true stress and strain cannot be properly estimated in micro-pillar compression [57], the intrinsic hardening is actually inaccessible. A larger "apparent hardening rate" in {100}-slip samples may stem from an enhanced dislocation storage ability, or from a different deformation pattern giving a distorted morphology. In section 4.4, we will quantify the differences between these two slip modes in terms of wildness and scaling exponent.

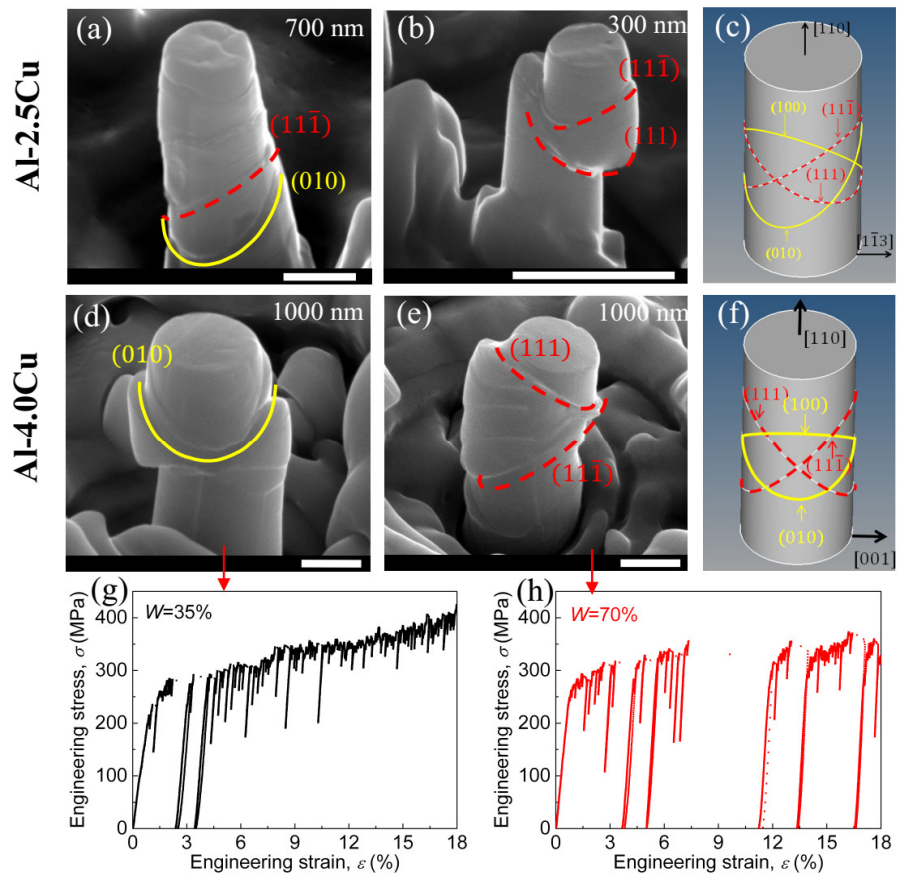

**Figure 4. Comparisons of deformation morphologies and engineering stress-strain curves for micro-pillars with {111}-slip and {111}+{100}-slip.** SEM images of Al-2.5Cu micro-pillars with diameters of $L = 700$ nm **(a)** and 300 nm **(b)**, and Al-4.0Cu micro-pillars with a diameter of $L = 1000$ nm **(d)** and **(e)**. Sketches in **(c)** and **(f)** represent the shapes of slip traces corresponding to the observed directions in Al-2.5Cu and Al-4.0Cu alloys respectively. The

Miller indices of coarse slip traces in SEM images are marked according to (c) and (f). **(g)** and **(h)** are the corresponding stress-strain curves to show the differences of mechanical response for micro-pillars of the same size with (d) and without (e) {100}-slip traces. All the scale bars represent 500 nm.

*4.3 Strength of micro-pillars*

In micro-pillar compression, the strength of samples can be characterized by the flow stress at a given strain value, typically between 2% and 10% [4,5], or, possibly, by the stress at the first detected strain burst [61]. However, our micro-pillars cover the wild and mild regimes. In the mild regime, such as the Al-2.5Cu micro-pillars with 4000 nm diameter (Fig. 2d), strain bursts are, by definition of wildness, rare and limited in size. In this case, the first detected burst is observed for large strains, far beyond the elastic-plastic transition. In other words, a yield criterion based on the detection of the 1st burst would depend on the degree of wildness itself. To minimize the effect of "apparent strain hardening", as well as problems in the estimation of stress and strain at large deformation, we simply use the *engineering* yield criterion of the stress at 0.2% plastic strain by following the works in Ref. [62-64]. Fig. 5a shows the critical resolved shear stress $\tau_{yield}$ on {111}-plane, calculated from this 0.2% offset stress multiplied by the Schmid factor (0.408 for [110]-orientation), as a function of external size *L*. The data for pure Al is taken from our earlier work [17]. On Fig. 5a, the stronger materials exhibit a weaker size effect, consistent with an enhanced role of disorder (i.e. external-size independent) on strength [36,37,65].

It should be noticed that the differences of strength between Al-Cu alloys and pure Al counterparts vanish as decreasing the sample size to 500 nm. This is shown on Fig. 5b where the strength gain, $\Delta\tau_{yield} = \Delta\tau_{yield}^{Al-Cu} - \Delta\tau_{yield}^{Al}$, is represented as a function of the external size *L*. $\Delta\tau_{yield}$ remains almost constant at large *L*, but decreases towards zero at a critical diameter. Such transition has been investigated in irradiated Cu micro-pillars [66], and attributed to a vanishing of dislocation-disorder interactions in small samples. A detailed analysis of strength is presented in section 5.2.

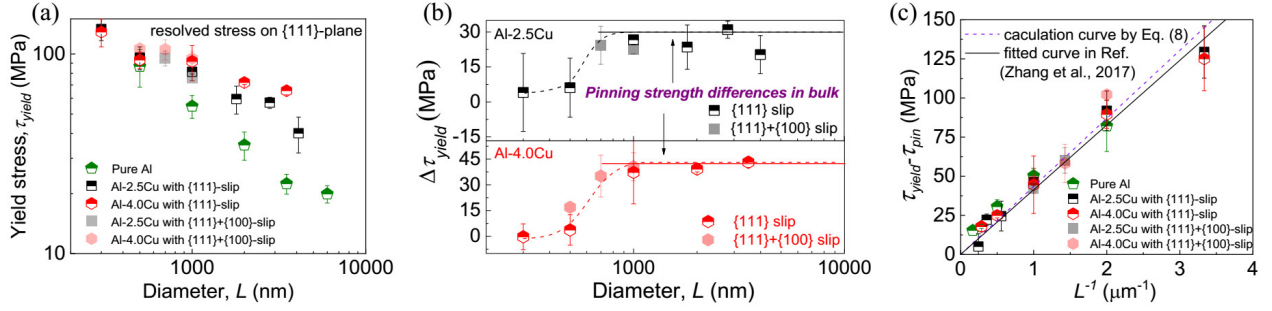

**Figure 5. External size dependent strength.** (a) Experimental results of the $L$-dependent yield stress (0.2% offset stress multiplied by Schmid factor) for Al-Cu alloys on {111}-plane, together with the data of pure Al taken from Ref. [17]. (b) Strength increment in Al-Cu alloys relatively to pure Al as a function of $L$. Below a critical diameter, the strengthening effect of precipitates vanishes. (c) The effective shear stress to activate truncated dislocation source, $\tau_{yield} - \tau_{pin}$, as a function of external size. A universal relation $\tau_{yield} - \tau_{pin} \sim 1/L$ can be identified. The black solid line is taken from Fig. 11 of Ref. [17] and the purple dash line is calculated by Eq. (8).

## *4.4 Statistics of fluctuations*

### *4.4.1 Methodology*

Determining the nature and origin of stochastic fluctuations in plasticity first relies on the quantification of their statistical properties [17]. Recently, we proposed a data mining procedure, based on the definition of an "avalanche" as a plastic process characterized by a dissipation rate much greater than the imposed loading rate, allowing to extract avalanches from smooth flow through an objectively determined threshold (see section 4.1 of Ref. [17] for details). The size of a burst is quantified by the axial plastic displacement released as a result of dislocation motion: $X = D_e - D_s + (F_s - F_e)/K_p$, where $D_s$, $F_s$, $D_e$ and $F_e$ are the displacement and force when a detected displacement jump starts and ends, respectively, and $K_p$ is the stiffness of the sample [17]. Theoretically, $X$ scales with the cumulative area swept by the dislocations involved in the avalanche, $\Delta A$, between the two elastic states $((D_s, F_s)$ and $(D_e, F_e))$: $X \sim b(\Delta A/A)$ [58], where $A$ is the area of the sample cross section. Once these bursts are extracted, a robust maximum likelihood methodology [67] is used to estimate the lower bond $X_{min}$ to power law scaling, and the scaling exponent $\kappa$. The fraction of plastic deformation accommodated through power law distributed bursts ($X > X_{min}$) defines the wildness $W$. More details can be found in our recent work (section 4.1 and SM of Ref. [17]).

*4.4.2 Effects of external size and disorder on wildness*

The statistical results of plastic fluctuations as a function of external size $L$ are presented in Fig. 6, in terms of wildness $W$ and scaling exponent $\kappa$. The trend of "dirtier is milder" is recovered for these two Al-Cu alloys, with a milder deformation (small $W$) in Al-4.0Cu compared with Al-2.5Cu micro-pillars (Fig. 6a), which we interpret as a reduction of the "mechanical temperature" of the system when increasing the pinning strength of the obstacles [17,32]. More specifically, the pinning effect acting on dislocations frustrates the long-range elastic mutual interactions [38] and facilitates short-range reactions as well as deformation isotropy (see section 4.2), thus driving plasticity towards a mild regime [17]. However, another typical effect of sub-$\mu$m plasticity, expressed as "smaller is wilder", is broken down at an intermediate diameter range: the wildness of the micro-pillars exhibiting {100}-slip traces drop from the general trend (Fig. 6a), thus confirming the qualitative statement in section 4.1. This anomalous behavior is interesting, as it may suggest a methodology to mitigate the detrimental effect of "smaller is wilder" in NEMS by introducing precipitates with a size comparable to device dimensions.

Besides the wildness $W$, the distributions of displacement burst sizes (compiled over the entire loading process) are also given in Fig. 6. For small samples (e.g., $L$ = 500 nm on Fig. 6c and d), the exponents $\kappa$ are slightly above the mean field theory [25,68] and DDD simulations [19,34,69,70] prediction, $\kappa$~1.5, corresponding to essentially power-law distributed fluctuations and a maximum wildness ($W \rightarrow 1$), while $\kappa$ is observed to increase significantly upon increasing the sample size, in association with a decreasing wildness. We noted above that micro-pillars with {100}-slip traces exhibited a lower wildness than the ones with {111}-slip only. This anomalous behavior is expressed also in term of scaling exponent, showing a larger $\kappa$ for the 700 nm Al-2.5Cu micro-pillars with {100}-slip than for the 1000 and 1800 nm micro-pillars of the same material with {111}-slip only (Fig. 6c). This association between larger scaling exponents $\kappa$ and {100}-slip is also observed for Al-4.0Cu micro-pillars (Fig. 6d).

We have shown previously from tests on pure Al and nm-sized disorder-reinforced Al alloys [17]

that the combined effects of external size and disorder on plastic intermittency manifested themselves through a unique (material independent) mapping between the power law exponent and wildness, $W(\kappa)$ (see the purple points in Fig. 6b). Here we show that this "universal" relation holds for the $\mu m$-sized precipitate reinforced Al-Cu alloys (Fig. 6b). Quite remarkably, it even holds for the micro-pillars with {100}-slip traces where different deformation mechanisms are believed to operate (solid semi-transparent symbols in Fig. 6b).

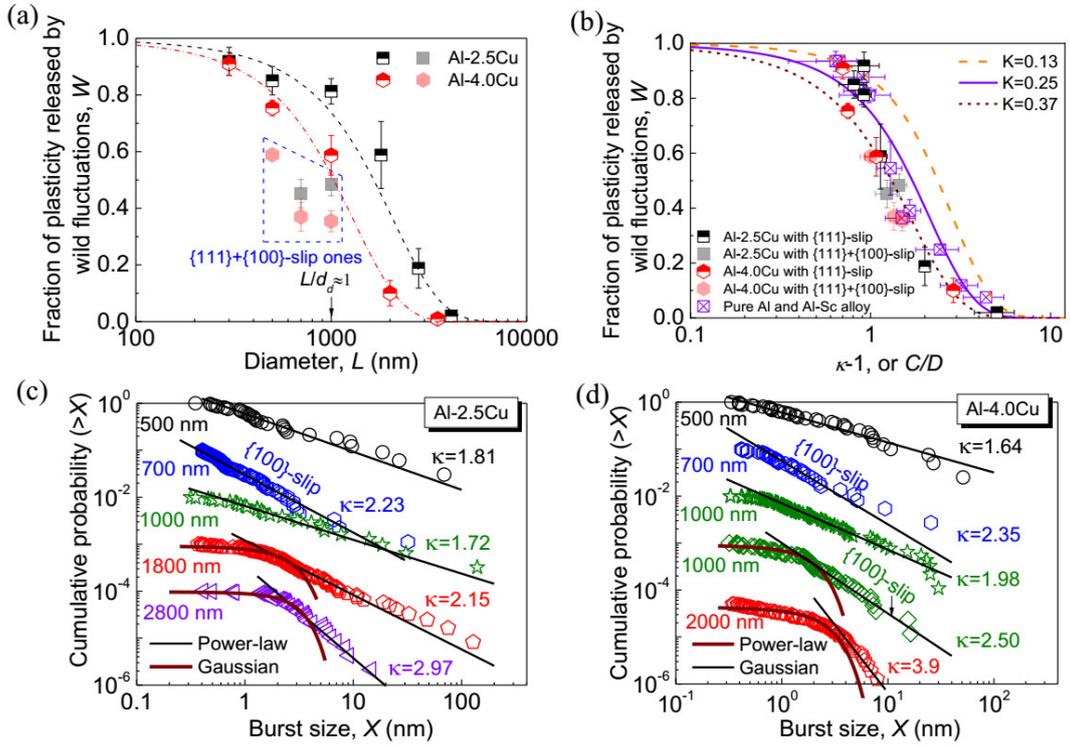

**Figure 6. Statistics of intermittent plasticity. (a)** The external size dependent wildness of Al-Cu alloys. The solid semi-transparent symbols represent the micro-pillars showing {111}+{100}-slip traces. **(b)** The universal relation between wildness $W$ and power law exponent $\kappa$-1. The data for pure Al and Al-Sc alloys are taken from Ref. [17]. **(c)** and **(d)** Typical strain burst distributions in Al-2.5Cu and Al-4.0Cu alloys respectively, where the samples showing {100}-slip traces are labelled. The burst distributions of samples with 1000 nm diameter in (d) correspond to the micro-pillars of Fig. 4d and e.

## 5 Discussion

### *5.1 Length scales vs regimes of plastic deformation*

The dependence of the mechanisms of plasticity on external size $L$ in single crystals can be speculated from a rough length scale analysis, comparing $L$ with the microstructural length scales of dislocation networks, $l_{dn}$, and disorder arrays, $l_{da}$. For our case, the characteristic length scale of disorder arrays $l_{da}$ includes the diameter of plate-like precipitate, $d_d$, as well as the spacing between precipitates ($\lambda$ and $l_c$, see section 3.1).

In the upper limit, $L \gg l_{dn}, l_{da}$, the samples are large enough to be representative volume elements (RVE) for the disorder and the dislocation network of the corresponding bulk alloys. Hence, we expect a plastic behavior controlled by the internal structure and characterized by an external size independent strength and mild deformation. The strengthening mechanisms of plate-like precipitates in this regime have been recently studied by micro-pillar compression [71]. In the lower limit, $L \ll l_{dn}, l_{da}$, the samples actually degrade to their pure microcrystal counterparts, with an absence of dislocation-disorder interactions before dislocation annihilation at free surfaces (schematically shown on Fig. 7c). This is exactly the case in our small samples, as shown by a vanishing strength difference between Al-Cu alloys and pure Al when $L \leq$ 500 nm (Fig. 5b). Between these two extreme cases, a complex regime emerges from a combination of external size and internal disorder effects. The case $l_{da} \ll L < l_{dn}$, where the sample can only be considered as a RVE for the disorders (not for the dislocation network), was considered in our previous work [17]. In this regime, the sample strength can be considered as a superposition of disorder strengthening and dislocation activation/nucleation stress [5,72], and the plastic fluctuations are controlled by the competition between the external size $L$ and a disorder related length scale $l$ through the parameter $R = L/l$ [17].

However, in case of $l_{da} \sim L < l_{dn}$, a strong interference between the external size and the disorder related length scales is expected, as for the present Al-Cu alloys. Detailed investigations in this regime are still lacking. We can speculate that the dislocations can bypass precipitates when $L/d_d >$ 1 (Fig. 7a), but would pile-up against the precipitate/matrix interface if the precipitates cut off the entire micro-pillar (Fig. 7b). However, beyond this simple reasoning, to what extent the mechanical properties in this regime fit into the previous frameworks, and what are the plastic mechanisms specific to these Al-Cu alloys, remain open questions. In what follows, we will first extract the features common to all alloys and length scales, and then discuss the specific plastic mechanisms of the Al-Cu alloys in the regime $l_{da} \sim L < l_{dn}$.

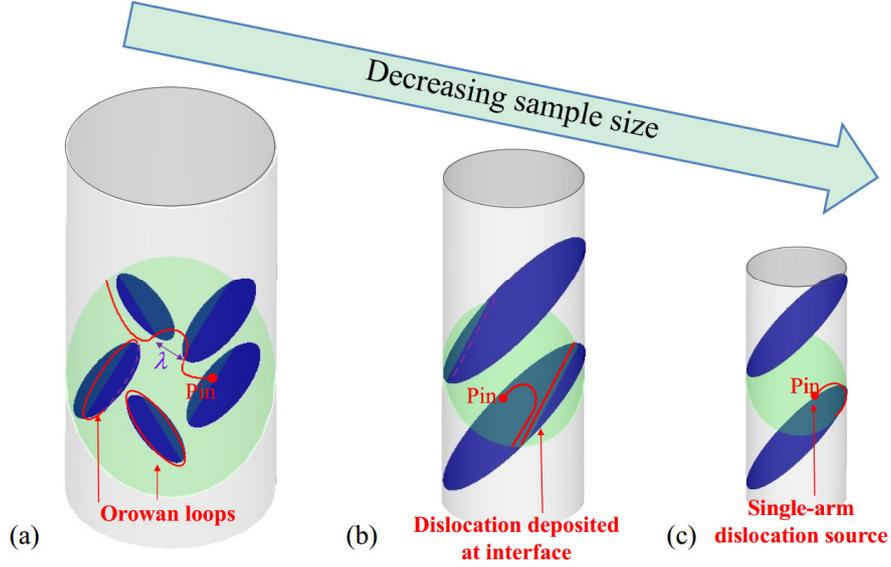

**Figure 7. Deformation mechanisms of Al-Cu micro-pillars. (a)** For large sample sizes, the precipitates are surrounded by the Al-matrix, thus the mobile dislocations can bypass precipitates through bulk-like Orowan mechanism. **(b)** For samples of intermediate size, the plate-like precipitates cut off the entire micro-pillar, forming a multilayer-like microstructure. In this case, the dislocations pile-up against the precipitate/matrix interface instead of bypassing precipitates. **(c)** For small samples, dislocations can run across the micro-pillar within the interval between two precipitates, without interacting with them.

## *5.2 Universal features in plastic fluctuations and yielding*

### *5.2.1 Dislocation-disorder interactions*

Analyzing the strength of Al-Cu micro-pillars can shed light on the dislocation-disorder interactions that play a crucial role in plastic avalanches. For $\mu m$ to sub-$\mu m$ pillars, their strength has been considered to be a superposition of the pinning strength of disorder and the activation stress of the weakest single-arm dislocation source (SAS) [10,73-76]:

$$\tau_{yield} = \tau_{pin} + \frac{\alpha G b}{l_s(\rho,L)} \qquad (5)$$

where $l_s$ is the length of the weakest SAS, and $\alpha$ is a geometrical constant of order of unity. To estimate $\tau_{pin}$ for micro-pillars, we compare on Fig. 5b the strength difference between Al-Cu and pure Al micro-pillars, $\Delta\tau_{yield}$, with the corresponding difference for bulk samples, $\Delta\tau_{bulk}$. From Eq. (5), we expect $\Delta\tau_{yield} = \tau_{yield}^{Al-Cu} - \tau_{yield}^{Al} = \tau_{pin}^{Al-Cu} - \tau_{pin}^{Al} = \Delta\tau_{bulk}$, which is recovered for sample sizes $L \geq 700$ nm. This means that the strengthening mechanisms acting in these samples are essentially unchanged compared with bulk samples, i.e. $\tau_{pin}$ can be safely estimated from Eq. (4). On the other hand, when $L \leq 500$ nm, as shown in section 4.3 and discussed above, dislocation-precipitate interactions vanish and the samples behave as pure Al. Within this diameter range, we can therefore consider $\tau_{pin}^{Al-Cu} = \tau_{pin}^{bulk\ Al}$.

*5.2.2 Wild-to-mild transition*

The typical effect of sub-$\mu$m, expressed as "smaller is wilder", is quantified in Fig. 6a. The enhanced role of the free surface as decreasing $L$ reduces the possibility of mutual dislocation reactions, leading to an absence of self-induced dislocation microstructure that suppresses plastic instability in large samples [32]. In addition, the increasing external stress can break the dislocation entanglements, resulting in pronounced dislocation avalanches (see the model and the experimental relations between burst size and external stress in [17,21,28]). The wild-to-mild transition can be shifted to small size by introducing disorders.

In our previous work [17], we translated the pinning strength of disorders into a characteristic length scale $l = Gb/\tau_{pin}$ characterizing the competition between disorder-dislocation interactions and long-range elastic mutual interactions between dislocations. *l* represents a distance at which the long-range elastic stress associated with a dislocation becomes similar to the pinning stress of disorder. This way, we found that a single non-dimensional parameter $R = L/l$ unifies the transition from wild to mild fluctuations for different materials. In the present study, using the $\tau_{pin}$ values for micro-pillars as explained in the above section, we are still able to gather the wildness values of {111}-slip Al-2.5Cu and Al-4.0Cu samples onto a single curve (Fig. 8a), and the overall shape (sigmoidal) of the wild-to-mild transition remains unchanged. This suggests that the main features of the wild-to-mild transition in the Al-Cu samples can still be captured by our previous framework, meaning that the effect of pinning strength on the long-range elastic interactions between dislocations remains a major factor on preventing intermittency in the Al-Cu alloys.

Since we considered above only the slip resistance on {111}-planes, it is not surprising to observe that {100}-slip micro-pillars deviate from the $W(R)$ curve of Al-Cu alloys (Fig. 8b). Intro-ducing the Peierls stress of the dislocation glide along the coherent precipitate/matrix interface, calculated to be ~ 80 MPa (see the companion paper [45] for details), into the relation $l = Gb/\tau_{pin}$ to estimate $R$, the data points of {111}+{100}-slip samples move onto the sigmoidal line (Fig. 8b), indicating a contribution from this non-trivial slip mode to the inhibition of plastic fluctuations.

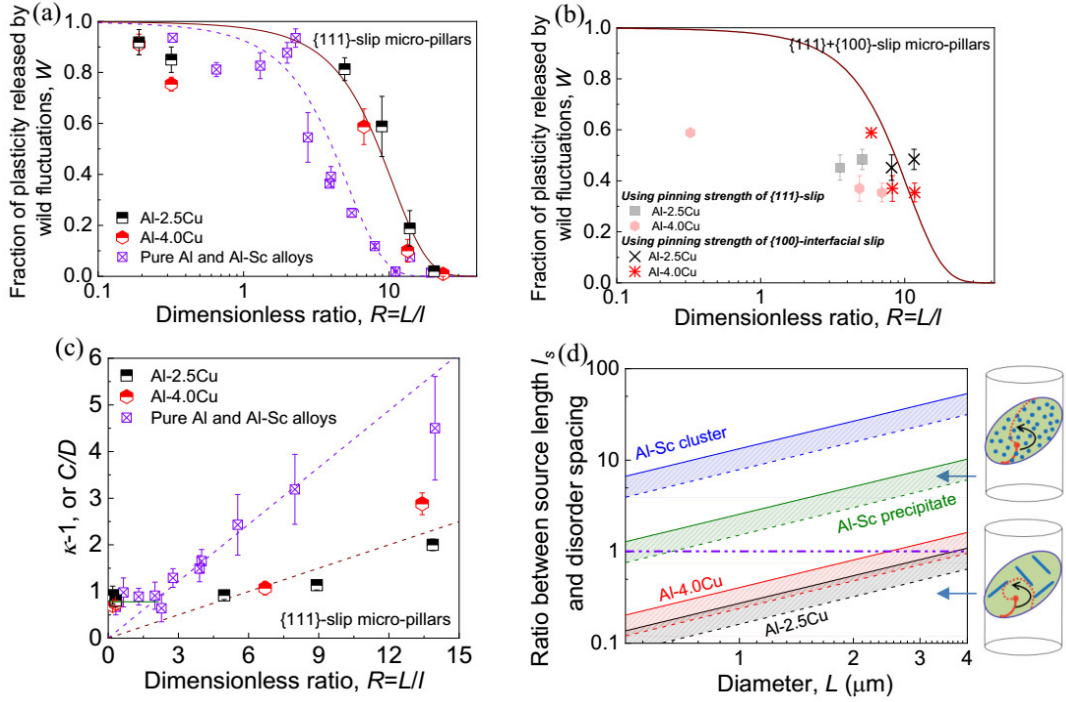

**Figure 8. Statistics of intermittent plasticity as a function of $R$ and dislocation-disorder interaction map. (a)** The wildness $W$ of {111}-slip micro-pillars as a function of non-dimensional ratio $R$. The sigmoidal curves are calculated from the model, using the universal relation $W(C/D)$ in Eq. (6) and the linear relation $C/D \sim L/l$ in (c). **(b)** The wildness $W$ of {111}+{100}-slip micro-pillars as a function of $R$. The pinning strength of dislocations slipping along both {111}-plane of matrix and {100}-interface are used to estimate $R$. The solid curve is as the same as the one in (a). **(c)** The relation between the power law exponent $C/D = \kappa - 1$ and $R$ for {111}-slip micro-pillars. **(d)** The ratio between the length of dislocation sources $l_s$ and the disorder spacing (λ for solution clusters and $nm$-sized precipitates, and $l_c$ for $\mu m$-sized precipitates) as a function of external size $L$. Both initial and stable dislocation sources are considered, giving the ratio range in (d). The sketches in (d) are to illustrate the differences of pinning landscapes in the two regimes. The data of pure Al and Al-Sc alloys are taken for Ref. [17].

*5.2.3 The universal relation $W(\kappa)$*

Instead of mean-field theory arguing for a universal landscape of intermittent plasticity [25], it has been revealed that the associated statistical properties are material- and size-dependent, with varying scaling exponent $\kappa$ and wildness $W$ [17,32,77]. Note that the loading method in these experiments remain the same, and the varying $\kappa$ should not be an effect of machine stiffness analyzed in [78]. On the other hand, we found that the relationship between these two variables, $W(\kappa)$, in pure

Al and nano-sized disorder reinforced Al alloys, is "universal" (material-independent) [17]. In the present study, we extend this relation to the Al-Cu alloys containing $\mu m$-sized precipitates (Fig. 6b). This is quite remarkable, as it suggests that this relation not only holds for materials containing large-sized precipitates, but also for samples showing totally different deformation mechanisms ({100}-slip and distorted morphology). In other words, this indicates that the $W(\kappa)$ relationship may be truly universal, even does not depend on microscopic plastic mechanisms.

This is in full agreement with our previously proposed stochastic model for the evolution of mobile dislocation density $\rho_m$ that gives [17,32]:

$$W = 1 - \frac{\Gamma\left(\frac{C}{D}, log\left(\frac{1}{K}\right)\right)}{\Gamma\left(\frac{C}{D}, 0\right)} \tag{6}$$

where $C$ is the rate of mutual annihilation/immobilization of the dislocation pairs; $D$ is the intensity of a multiplicative mechanical noise that describes fluctuations experienced by a representative volume due to interactions with the rest of the system, which can also be seen as the mechanical temperature of the system; $\Gamma(z,x) = \int_x^\infty t^{z-1} e^{-t}\, dt$, and $K$ is a parameter setting a threshold to distinguish the power law tail in the model. In this model, the ratio $C/D$ is linked to the power law exponent, $C/D = \kappa - 1$ [32]. Although $C/D = \kappa - 1$ and $W$ depend on external size and plastic mechanisms, Eq. (6) predicts that $W$ depends solely on $C/D$, as shown on Fig. 6b, whatever the material, and is independent of microscopic plastic mechanisms. Note that Eq. (6) weekly depend on parameter $K$ [17]. Based on the universal relation $W(C/D)$ in Eq. (6) and the scaling $C/D \sim L/l$ in {111}-slip Al-Cu alloys identified on Fig. 8c, the wild-to-mild transition on Fig. 8a can be recovered.

*5.2.4 Yield stress scaling*

As discussed in section 5.2.1, micro-pillar yield stress is considered to be a superposition of the pinning strength of disorder and the activation stress of the weakest SAS, $\tau_{yield} = \tau_{pin} + \frac{\alpha G b}{l_s(\rho,L)}$ (Eq. (5)), where $l_s$ is a function of the dislocation density $\rho$ and external size $L$ [10,63,79,80]. An expression for the length of the weakest SAS at the *onset* of plastic flow was proposed from 3D discrete dislocation dynamic (DDD) simulations of pure single crystals [79], for $\rho L < 10^7$ m$^{-1}$, i.e. when considering an initial dislocation density of order $10^{12}$ m$^{-2}$, for sample sizes typically below few μm:

$$l_s = \frac{bL\sqrt{\rho}}{\eta} \tag{7}$$

where the parameter $\eta = 1.76 \times 10^{-3}$ is a dimensionless constant. Combining Eq. (5) and (7) we get:

$$\tau_{yield} - \tau_{pin} = \frac{\alpha G \eta}{\sqrt{\rho}} \frac{1}{L} \tag{8}$$

This scaling $\tau_{yield} - \tau_{pin} \sim 1/L$ is consistent with a re-evaluation of a large set of published experimental data [65], and has been verified in our pure Al and Al alloys in Ref. [17]. Using the pinning strengths obtained in section 5.2.1, a remarkable convergence of data, including pure Al and Al alloys previously studied [17] as well as the present results, towards a single master curve is obtained (Fig. 5c). Taking an initial dislocation density of $\rho = 10^{12}$ m$^{-2}$ for well annealed samples and $\alpha = 1$, Eq. (8) fits the experimental data very well (dashed line on Fig. 5c).

The universality of Eq. (8), for pure Al to Al alloys, indicates that $l_s(\rho, L)$ in Eq. (7) can be extended to Al alloys at plastic yield, i.e., the $l_s(\rho, L)$ at the *onset of plastic flow* only depends on dislocation density and sample size, not on disorder. Moreover, on Fig. 5c we only considered the resolved shear stress on {111}-planes, even for samples with {111}+{100} slip traces. Still, these {111}+{100}-slip samples fit the universal relation Eq. (8), indicating that the yield stress in these samples should be essentially controlled by {111}-slip, in contrast with a wildness controlled (reduced) by {100}-slip (see section 4.4.2). Indeed, if the initial stage of plasticity is controlled by the weakest slip system, upon increasing the strain further, other mechanisms such as {100}-slip take place and reduce the wildness. This is consistent with the observation that, in these samples, the jerkiness of stress-strain curves decreases with increasing strain (Fig. 4g).

## 5.3 Specific mechanisms in Al-Cu micro-pillars

### 5.3.1 Disorder effect on taming intermittency

In section 5.2.2, translating the pinning strength into an internal length scale, $l = Gb/\tau_{pin}$, we found that the dimensionless ratio $R = L/l$ unifies the wild-to-mild transitions of Al-2.5Cu and Al-4.0Cu alloys. However, compared with the previous Al alloys containing nano-sized disorders [17] (purple points on Fig 8a), the sigmoidal transition is shifted for the present alloys towards larger $R$-values. Actually, for materials with strong lattice resistance or dense nano-sized disorders, the pinning effect works on a scale of atomic distance, or of disorder spacing, much smaller than the average length of mobile dislocation segments. In this case, the mobile dislocation segments must overcome this pinning effect to move. However, this mean-field pinning picture is questionable for Al-Cu micro-pillars, where the precipitates spacing is commensurate with the sample size, hence with the length of dislocation segments. To explore this situation, we take the SAS length $l_s$ as a characteristic length of mobile dislocation segments, considering that plasticity is largely associated with the activation and shutdown of SAS in the source-truncation regime [9], and compare $l_s$ with the precipitate spacing determined in section 3.1.

Eq. (7) gives $l_s$ at the onset of plastic flow. This source length should evolve with deformation. Cui et.al [81] gives an expression of the stable SAS length at the flow stage, for micro-pillars with a typical initial dislocation density obtained in FIB-milled specimens:

$$l_s = \frac{L}{4\cos^2(\beta/2)} \tag{9}$$

where $\beta$ is the angle between the normal to the slip plane and the loading axis. This expression gives $l_s = 0.317\ L$ for the [100]-oriented micro-pillars, which has been validated by the statistic measurements of $l_s$ in their DDD simulations, and by the work of Ryu et al. [80]. For the [110]-oriented micro-pillars of the present study, Eq. (9) gives $l_s = 0.254\ L$. The characteristic length of dislocation segments, both at the yield (Eq. (7)) and flowing (Eq. (9)) stages, were compared with the largest length scale characterizing precipitate spacing, which is $l_c$ for plate-like precipitates, and $\lambda$ for nanometer sized disorder reinforced alloys (see section 3.1). Fig. 8d shows that, in contrast with the Al alloys micro-pillars of Ref. [17], the source lengths in the present Al-Cu micro-pillars are smaller than the precipitate spacing, $l_s/l_c < 1$. This clearly means that the weakest dislocation sources can be activated and then move inside the plate-like precipitate array before getting pinned (see the sketch for Al-Cu alloy in Fig. 8d). This breakdown of the mean-field pinning picture, a typical feature of alloys containing plate-like precipitates of size comparable with the external size, seems to not affect the *global* properties, such as yield strength scaling (Fig. 5c). However, it makes the dislocation motion sensitive to small stress perturbations at *micro-scale* before getting pinned. In our model words, it increases the *local* "mechanical temperature" of the system [17,32], leading to wilder fluctuations when comparing with the mean-field pinning case (upper sketch on Fig. 8d) with a same $R$.

*5.3.2 Precipitate shearing*

As schematically shown in Fig. 7b, when the external size $L$ becomes comparable with the precipitate diameter $d_d$, the precipitates can cut off the micro-pillar, forming a multilayers-like structure. This structure, which can be seen in the TEM image of the longitudinal section of a micro-pillar with {100}-slip traces (Fig. 9a), prevents dislocations to bypass the obstacles. Consequently, dense dislocation networks are observed near the precipitates (Fig. 9b) as a result of dislocation pile-ups. The straight dislocations in Fig. 9c and d, with lengths much larger than the thickness of the foil, necessarily lie within the TEM foil plane (100). Therefore, these (100)-dislocations with zero Schmid factor are most likely to be the geometrically necessary dislocations (GNDs) generated to accommodate the strain gradient near the penetrating plate-like precipitate. These GNDs can also serve as obstacles to mobile dislocations, hence inhibiting the plastic intermittency ever further. In polycrystalline materials, pile-up-induced stress concentrations are eventually relaxed by the activation of dislocations in neighboring grains [82], or by the penetration of leading dislocations through grain boundaries [83]. In our case, we observed instead the shearing of precipitates.

The high-angle annular dark field image (Fig. 10a) and the energy dispersive spectrometer (EDS) of Cu atoms (Fig. 10b) show a broadened precipitate with a wavy and diffused interface at the top of 1000 nm Al-2.5Cu micro-pillar (marked on the top of the micro-pillar in Fig. 9a), in contrast with the straight and sharp interface of the undeformed precipitate (Fig. 1a and b). This precipitate broadening, a sign of the frequent shearing of precipitates by matrix dislocations, has also been observed in the aged Al-Cu alloys experiencing severe plastic deformation (average shear strain larger than 100%) [84,85]. The atomic steps at the precipitate interface (Fig. 10c) also supports the shearing mechanism proposed above. Moreover, the plasticity of intermetallic compound can be fundamentally different from that in Al matrix. A previous work showed that dislocation motion in $\theta$-Al$_2$Cu precipitate induces a local deviation of the composition from its stoichiometric ratio, and the lattice is highly distorted into a disordered-like structure [86]. Similarly, as a result of precipitate shearing by matrix dislocations, we also observed a highly disordered lattice structure in our deformed $\theta'$-Al$_2$Cu precipitate (see also the simulation result in Fig. 4d of the companion paper [45]). This precipitate shearing mechanism increases the overall pinning strength $\tau_{pin}$ at large strain for the high shear resistance (~1.06 GPa [53]), thus reducing the internal length scale $l = Gb/\tau_{pin}$. The combination of the abnormal {100} interfacial slip and precipitate shearing should be responsible for the sharp decrease of wildness in the micro-pillars containing penetrating $\theta'$-Al$_2$Cu plate-like precipitates (Fig. 6a).

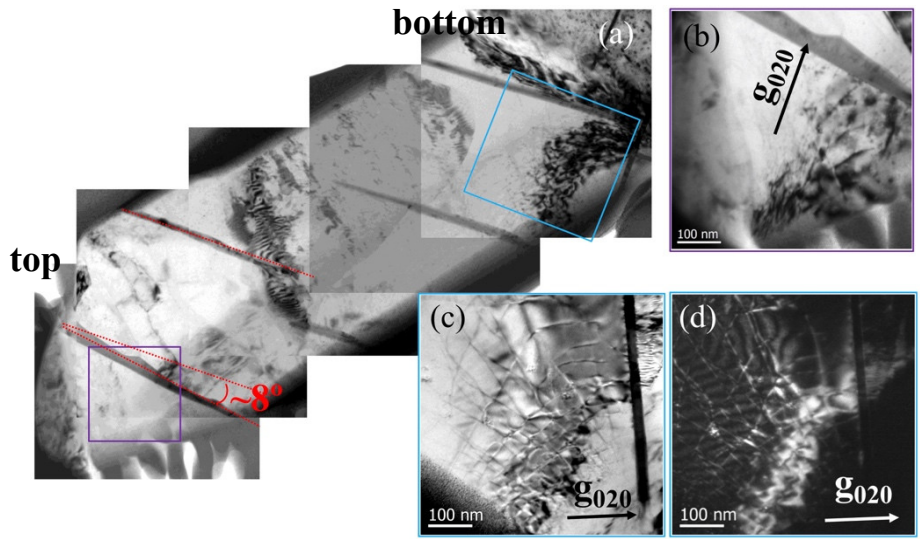

**Figure 9. TEM images of a deformed Al-4.0Cu micro-pillar with {100}-slip traces. (a)** The cross section of the micro-pillar observed along [001]-direction. A ~8° rotation for the marked precipitate indicates a strong strain gradient in this region. **(b)** and **(c)** are the bright-field TEM images of the regions marked respectively on the left and right part of (a). The irregular shape of $\theta'$-Al$_2$Cu precipitates can be seen in (b). **(d)** is the corresponding weak beam dark-field image of (c). The dislocation networks can be observed near the precipitate in (b), (c) and (d).

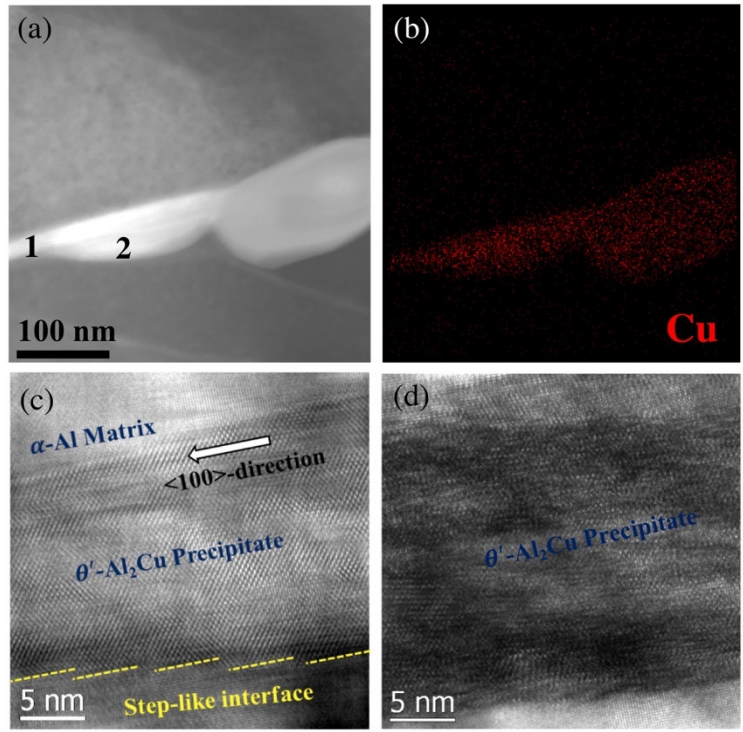

**Figure 10. Characterizations of $\theta'$-Al$_2$Cu precipitates in the deformed micro-pillar. (a)** and **(b)** are a HADDF image and its corresponding EDS Cu-mapping of a highly deformed $\theta'$-Al$_2$Cu precipitate marked at the top of the micro-pillar in Fig. 9a. **(c)** is the HRTEM image at the slight decomposition zone 1 marked in (a). **(d)** is the HRTEM taken from the severe decomposition zone 2 marked in (a). The absence of clear lattice periodicity can be found.

*5.3.3 Taming intermittency by using penetrating precipitates*

We showed above that plastic slip along {100}-planes and the shearing of precipitates due to dislocation pile-ups are the main deformation mechanisms when precipitates cut off the micro-pillars. In order to precise the role of these specific mechanisms in taming plastic intermittency, two [110]-oriented micro-pillars with square cross-section (~220×220 nm), one containing precipitates and the other not, were fabricated at the edge of an Al-2.5Cu foil and compressed in-situ under TEM (Fig. 11). For the sample without precipitate, the deformation behavior was, as expected, similar to that of pure Al samples with a similar size [17,47], with plastic strain highly localized on a single slip plane (Fig. 11b and c). The generated dislocations quickly slipped across the sample in an intermittent way, and wild deformation was observed (Fig. 11d). In contract, in the micro-pillar containing penetrating precipitates (Fig. 11e), the dislocations piled-up against the $\theta'$-precipitate (marked in Fig. 11f) upon gradually increasing the displacement. The configuration of dislocation pile-ups is similar to the TEM observation of Fig. 9b. Upon deforming the micro-pillar further, these pile-up dislocations sheared the $\theta'$-precipitate, resulting in a broadening of the precipitate (Fig. 11g). The {100}-slip steps can also be identified at the tip of the $\theta'$-precipitate (black arrows in Fig. 11g). These mechanisms tame intermittency and prevent the development of strain localization, inducing a homogenous diameter expansion (marked in Fig. 11g) and mild plasticity (Fig. 11h). Moreover, the penetrating $\theta'$-Al$_2$Cu plate-like precipitate can limit the space for collective dislocation motions, hence inhibiting the intermittent plasticity. These in-situ compression tests confirmed that the introduction of penetrating precipitates promotes strain homogeneity while taming plastic intermittency effectively.

In addition, we mention that the plastic flow of pure Al micro-pillar penetrated by a slanted grain boundary (GB) is more stable than its single crystal counterpart, due to the dominating role of GB gliding [87]. This is quite similar to our same sized micro-pillar penetrated by $\theta'$-Al$_2$Cu plate-like precipitates (Fig. 11), showing a suppression of jerkiness accompanied by the interfacial slipping. The difference is that the penetrating $\theta'$-Al$_2$Cu precipitates render an enhanced strain-hardening, rather than a strain-softening due to the grain boundary gliding [87], thus implying a potential advantage by using the penetrating plate-like intermetallic on stabilizing the plastic flow.

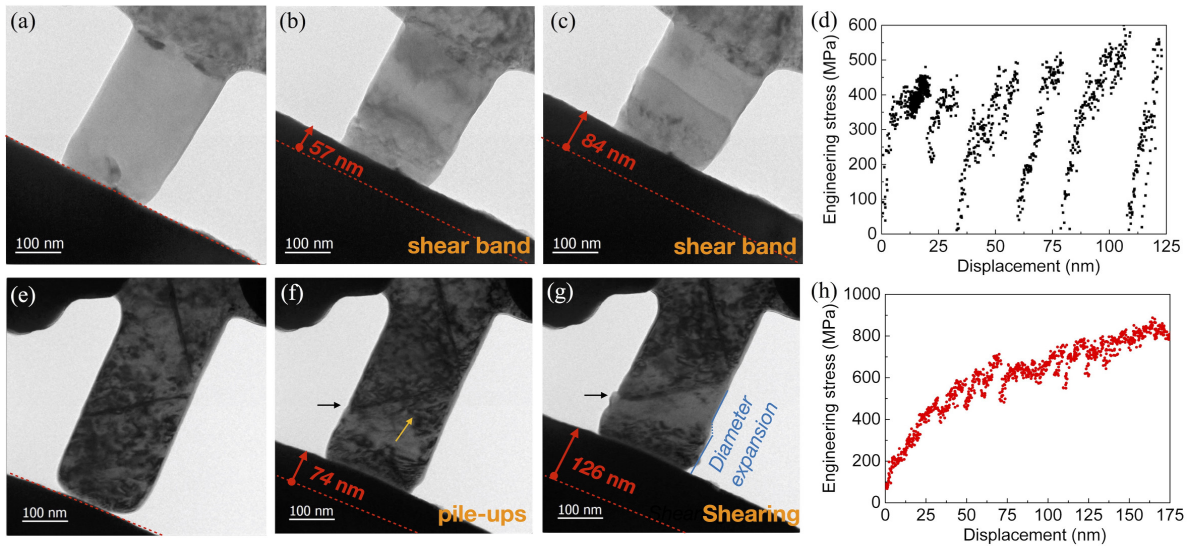

**Figure 11. In-situ observations of deformation behaviors of Al-2.5Cu micro-pillars. (a), (b) and (c)** TEM images of a micro-pillar without precipitates. The formation of a large slip band can be identified in (b) and (c). **(d)** Corresponding engineering stress-displacement curve. **(e), (f) and (g)** TEM images of a micro-pillar containing two orthorhombic precipitates. Dislocation pile-ups against precipitates, marked by a yellow arrow, can be observed in (f). **(g)** The pile-up dislocations marked in (f) sheared the precipitate, resulting in a precipitate broadening. The slip traces along the precipitates ({100}-plane) are pointed by black arrows in (f) and (g). **(h)** Corresponding engineering stress-displacement. The tip displacements are marked in the TEM images.

# 6 Conclusions

We performed micro-compression tests on [110]-oriented Al-Cu alloys micro-pillars containing plate-like $\theta'$-Al$_2$Cu precipitates. The microstructural length scales of precipitates are of the order of micrometer, commensurate with the external size $L$ of micro-pillars. Plastic fluctuations, yield strength and deformation mechanisms were investigated. Our results revealed that:

(1) Plastic fluctuations can be effectively inhibited by plate-like precipitates. Translating the pinning strength into an internal length scale $l = Gb/\tau_{pin}$, the dimensionless ratio $R = L/l$ unifies the wild-to-mild transition, showing that this inhibition of dislocation avalanches results essentially from the pinning effect of precipitates.

(2) In comparison with Al alloys containing dense nm-sized disorders, plastic deformation is wilder in the present Al-Cu alloys with the same pinning strength. We explained this phenomenon as a breakdown of the mean-field pinning picture, which is a specific feature of these Al-Cu alloys with an external size commensurate with precipitate spacing and size.

(3) A sharp decrease of wildness was observed when the plate-like precipitates cut off the entire micro-pillar. This observation suggests a new way to mitigate the "smaller is wilder" effect in the design of small devices.

(4) We observed precipitate shearing and nontrivial {100}-slip traces in Al-alloys over a specific diameter range. While the associated mechanisms giving rise to this unusual slip mode are detailed in our recent work, the high slip resistances of these specific mechanisms lead to the sharp decrease of wildness observed in this sample size range.

(5) The universal relation between wildness $W$ and the power-law exponent $\kappa$ of the size distribution of strain bursts was proved to be truly universal, regardless of the nature of the disorder (pure metal, solutes, precipitates), the ratio between the characteristic scales of disorder and the sample size, or even the nature of the underlying deformation mechanisms.

In summary, our study enlightens some generic laws of microscale plasticity common to various materials, while also addressing some specific deformation mechanisms in Al-Cu alloys. It also suggests a new approach to mitigate the "smaller is wilder" effect by plate-like precipitates. For a practical use, one should increase the precipitate density while reducing the variability of the precipitate size (e.g. by micro-alloying) in order to avoid the sample-to-sample scatter of the wildness in the regime with {100}-slip traces (Fig. 6a).

## Acknowledgements

This work was supported by the National Natural Science Foundation of China (Grant Nos. 51621063, 51722104, 51625103, 51790482, 51761135031 and 51571157), the National Key Research and Development Program of China (2017YFA0700701, 2017YFB0702301), the 111 Project 2.0 of China (BP2018008), and the China Postdoctoral Science Foundation (2019M653595). The financial support by the NSFC-ANR joint project SUMMIT (n° 17-CE08-0047, 51761135031) is sincerely acknowledged. We thank ShengWu Guo and YanHai Li for the TEM characterizations. Peng Zhang thanks the financial support from China Scholarship Council.